\documentclass[%
 aps,
 pra,
 reprint,
 twocolumn,
 superscriptaddress,
 showkeys, 
 floatfix
]{revtex4-2}

\usepackage[english]{babel}
\usepackage{graphicx}        

\usepackage{amsmath} 
\usepackage{amssymb} 
\usepackage{bm}              
\usepackage{physics} 

\usepackage{siunitx}         
\usepackage[
    colorlinks=true,       
    linkcolor=blue,         
    citecolor=magenta,      
    filecolor=cyan,         
    urlcolor=teal,
    bookmarks=true,      
    bookmarksopen=true,     
   breaklinks=true,         
]{hyperref}

\usepackage[table]{xcolor}
\begin{document}
\title{\textbf{Quantum-enhanced photocell based on GaN quantum dots}}
\author{Aditya Dev}
\thanks{These authors contributed equally to this work}
\affiliation{Department of Electronics and Communication Engineering, Indian Institute of Technology, Roorkee, Uttarakhand, 247667, India}
\author{Sumit Chaudhary}
\thanks{These authors contributed equally to this work}
\affiliation{School of Computation, Information and Technology, Technical University of Munich, Germany}
\author{Jay Reshamiya}
\affiliation{Department of Physics, Indian Institute of Technology, Roorkee, Uttarakhand, 247667, India}
\author{Abhishek Chakraborty}
\affiliation{Dept. of Physics and Astronomy, University of Rochester, NY, 14627, USA}
\author{Vishvendra Singh Poonia}
\email{Corresponding author: vishvendra@ece.iitr.ac.in}
\affiliation{Department of Electronics and Communication Engineering, Indian Institute of Technology, Roorkee, Uttarakhand, 247667, India}
\date{\today}

\begin{abstract}
    In this work, we propose an efficient quantum-enhanced solid-state photocell based on GaN quantum dots. We exploit the strong built-in electric field in GaN QDs and excitonic dipole-dipole coupling between adjacent QDs to break detailed balance, leading to enhanced device performance. This mechanism is significantly stronger than Fano interference, and our results demonstrate that such a photocell exhibits increased photovoltage and photocurrent compared to its classical counterparts. Numerical simulations further show that the efficiency remains positive and saturates at a finite value for multi-quantum dot systems. The proposed quantum photocell represents a promising step towards harnessing quantum effects in practical energy-harvesting devices.
\end{abstract}

\maketitle


\section{\label{sec:introduction}Introduction}
Photovoltaic or solar cells are one of the most promising options for clean energy. A major challenge in the design of solar cells is their low efficiency. It was shown by Shockley and Queisser~\cite{Shockley1961} that in junction-based solar cells, one of the causes of the limited efficiency is the loss of  electron-hole pairs due to radiative recombination before charge separation can occur to  extract useful work from them, a consequence of detailed balance. Quantum dot solar cells (QDSCs) have been  proposed as an alternative to p-i-n solar cells to overcome some of the other drawbacks of conventional photocell design.  Tandem cells with different bandgaps would utilise a larger part of the solar spectrum, increasing the overall  efficiency of a solar panel. Other approaches extract energy from hot charge carriers~\cite{Nozik2002}. 
We propose an alternative biologically inspired photocell design, which uses excitonic delocalization due to quantum interference  to  achieve enhanced efficiency. Originally, the idea of quantum photocell was proposed by Marlan O. Scully, where a theoretical analysis was presented on breaking the detailed balance~\cite{Scully2010}. Later, this notion was studied in the context of the photosynthetic complexes~\cite{Dorfman2013, Creatore2013}. In this paper, we present a solid-state quantum photocell based on GaN quantum dots (QDs). In the proposed design, excitonic dipole-dipole coupling in GaN QDs  creates the dark states, which practically reduces radiative recombination by breaking the detailed balance, resulting in increased efficiency over their classical counterparts. We have chosen GaN QDs as the material of choice for our quantum photocell, as GaN-based quantum dots have a wurtzite crystal structure in {AlGaN} bulk semiconductor, there is a strong built-in electric field~\cite{DeRinaldis2002}.  This field has contributions both a spontaneous polarization charge induced at the {GaN}/{AlGaN} interface, and a strain-induced piezoelectric field at the {GaN}/{AlGaN} interface~\cite{cingolani00}, and the latter dominates in the case of {GaN} quantum dots due to the lack of an inversion centre in the Wurtzite crystal structure. Due to the differing cell dimensions for {GaN} and {AlGaN} from the usual hexagonal structure, the spontaneous polarization is also an appreciable effect and is around the same order of magnitude and in the same direction as the piezoelectric field~\cite{andreev00}. It is also noted that the magnitude of the electric field is roughly equal both inside and immediately outside the dots, although there is a change in direction~\cite{DeRinaldis2002}.
This improves the lateral confinement of the exciton in these QDs. 
However, due to this separation of charges, there is a monotonic decrease in the excitonic binding energy as compared to a flat band~\cite{cingolani00}, which will aid the tunnelling of electrons in and out of the quantum dots. We analyze the proposed photocell for various configurations of quantum dots for the practically realizable parameters and examine where we get an advantage over its classical counterpart.

The manuscript is organized as follows: 
In \autoref{QPC}, we introduce the N-level system model for QDs used to design  the photocell, then describe the model with and without coupling. In \autoref{sec:tunneling_rates}, we calculate  the tunneling rates for charge carriers in and out of the dots. This is used to perform numerical simulations,  the results of which are presented in \autoref{sec:results} and \autoref{sec:contour_plots}.


\section{\label{QPC}The Quantum photocell model}
\begin{figure}[htbp]
	\centering
	\includegraphics[width=1\columnwidth]{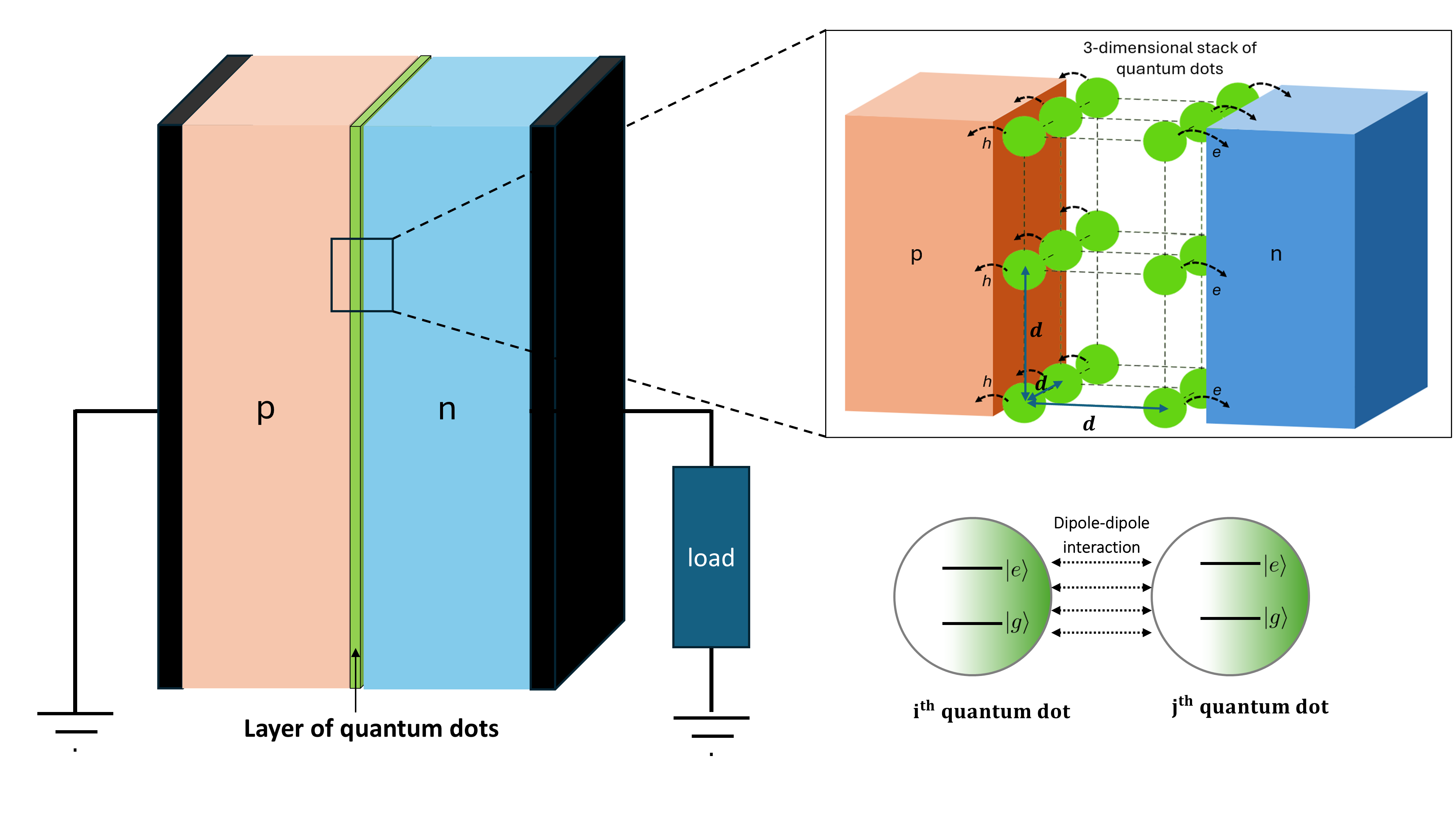}
	\caption{Schematic illustration of a QDSC featuring GaN quantum dots arranged in a three-dimensional lattice. Each quantum dot is separated from its nearest neighbors by a perpendicular QD spacing $d_{\perp}$, enabling dipole-dipole coupling within the array.\label{fig:schematic-qdsc}}
\end{figure}


\subsection{\label{sec:excitonic-dipole-dipole-coupling}The Coupled Quantum Dot System with Excitonic Dipole-Dipole Coupling}

\begin{figure*}[htbp]
\centering
\includegraphics[width=2.0\columnwidth]{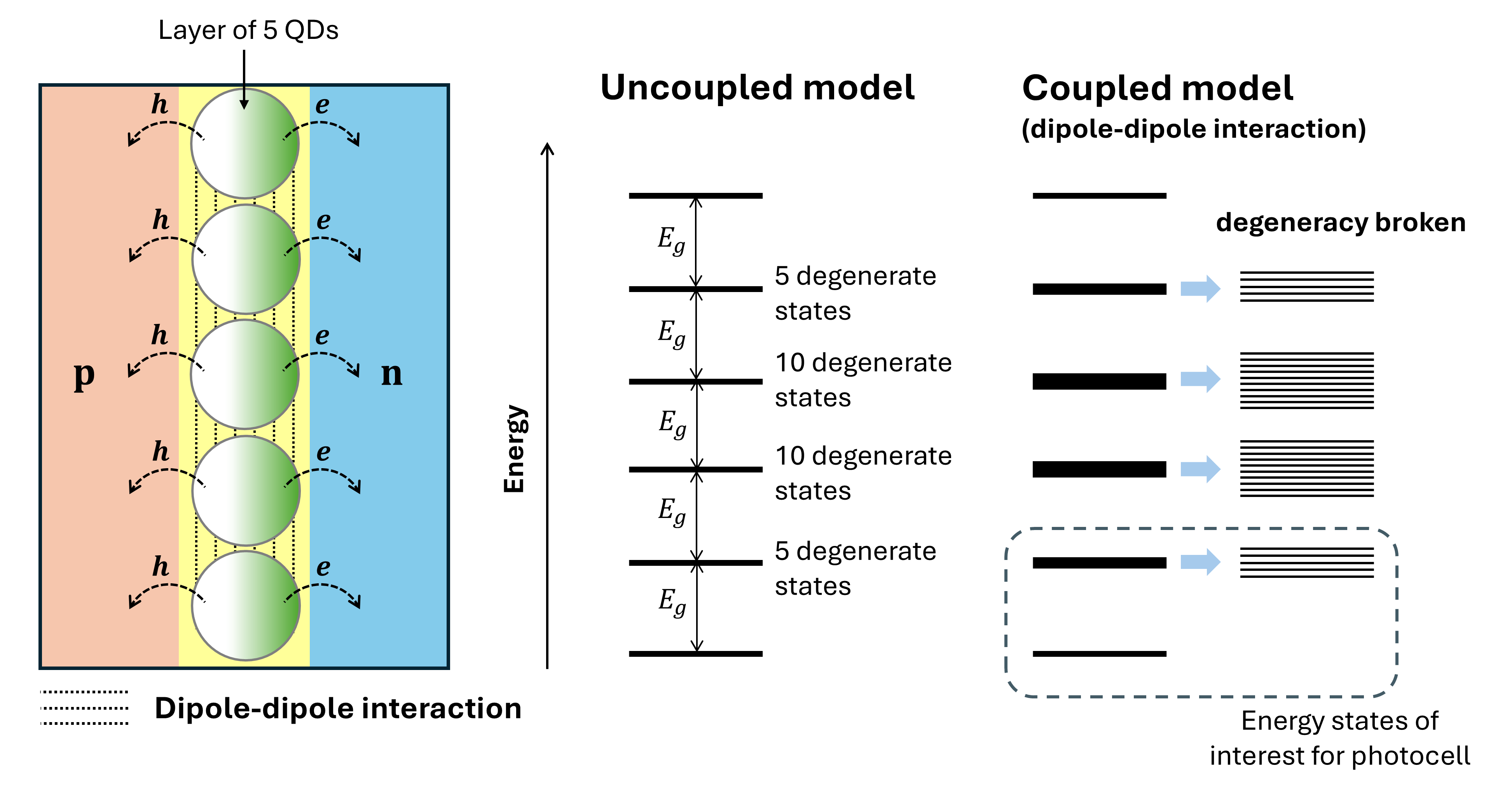}
\caption{Energy level diagram for a 5-quantum-dot (QD) system, comparing the uncoupled case (left) with the dipole-dipole coupled case (right). The interaction lifts the degeneracy of the first excited manifold, forming a new spectrum of collective states. \label{fig:energy_levels}}
\end{figure*}

Here, we present our model of a quantum dot-based photocell. The system is envisioned as a three-dimensional stack of quantum dots, where the perpendicular separation between adjacent dots is denoted by $d_\perp$, as illustrated in  ~\autoref{fig:schematic-qdsc}. As a key simplifying assumption, we consider the orientation of the quantum dots to be such that the excitonic dipole moment is always aligned along the x-axis, making all dipole moments parallel to each other. Each individual quantum dot (QD) is approximated as a two-level system (TLS) consisting of a ground state $\ket{g}$ (representing the absence of an exciton within the QD) and a single-exciton excited state $\ket{e}$. Let $E_{\mathrm{GaN}}$ denote the energy difference between these two states. Since excitonic states are bound electron-hole pairs, they possess a non-zero electric dipole moment. Consequently, when two QDs are sufficiently close, the exciton in one dot experiences the electric field induced by the excitonic dipole of the other, and vice versa. This dipole-dipole interaction (DDI) is the central mechanism of our model, leading to a modification of the collective system's excitonic energy levels, as presented in~\autoref{fig:energy_levels}.

\begin{figure}[htbp]
	\label{fig:dd-interaction}
	\centering	\includegraphics[width=1\columnwidth]{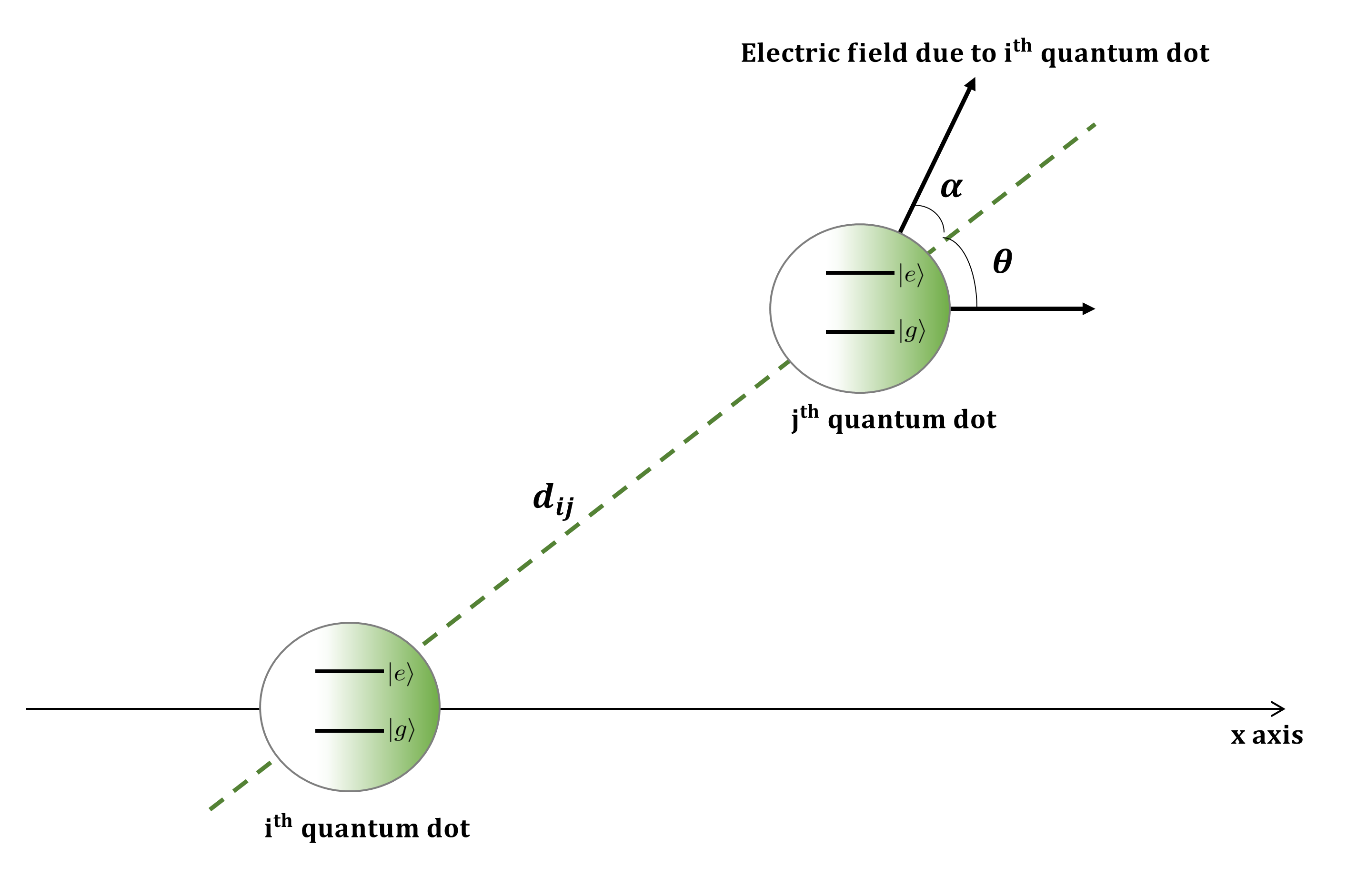}
	\caption{Geometric representation of the dipole-dipole interaction between two quantum dots, $i$ and $j$. The interaction strength $J_{ij}$ depends on the interdot distance $d_{ij}$ and the angles $\theta$ and $\alpha$, which define the relative orientation of the two dipoles.}
\end{figure}

For two excitonic dipoles \(\va*{\mu}_1\) and \(\va*{\mu}_2\) at position vectors \(\va*{r}_1\) and \(\va*{r}_2\) respectively. The electric fields generated at position vector $\vec{r}$ by these dipoles are \(\va{E}_1(\va*{r})\) and \(\va{E}_2(\va*{r})\). The interaction energy is given by the potential energy of one dipole in the electric field generated by the other:
\begin{equation}
\label{eq:dip-int-pot-energy}
\Delta E = -\va*{\mu}_2\cdot\va{E}_1(\va*{r}_{21}) = -\va*{\mu}_1\cdot\va{E}_2(\va*{r}_{12})
\end{equation}
where \(\abs{\va*{r}_{21}} = \abs{-\va*{r}_{12}} = \abs{\va*{r}_2 - \va*{r}_1} \equiv d_{\perp}\) equal to the perpendicular QD spacing. For the specific case relevant to our stacked geometry i.e. both dipoles are parallel and equal in magnitude (\(\va*{\mu}_1 =  \va*{\mu}_2 = \va*{\mu}\)), the electric field due to a dipole at a point on the perpendicular bisector plane of the other if give by
\begin{equation}
\label{eq:dip-per-field}
\va{E}_{\mathrm{dip}}^{\perp}(\va*{r}) = -\frac{1}{4\pi\epsilon} \frac{\va*{\mu}}{r^3}
\end{equation}
this results in a characteristic coupling strength between two adjacent QDs
\begin{equation}
\label{eq:dip-energy-shift}
J_0 \equiv\frac{|{\vec{\mu}}|^2}{4\pi\epsilon}
\end{equation}
We can construct a configuration such that the interaction is perfectly \emph{coherent} and thus no energy leaks out from the system into the radiation field. This is done by ensuring that the interdot spacing $d_{\perp} \ll 1/k$ where $k$ is the wavevector associated with the transition in the dots. To find the new excitonic spectrum for a multi-QD system, we must diagonalize the total Hamiltonian, which includes the DDI:
\begin{equation}
\label{eq:dipole_tls_ham}
H = H_{0} + H_{I}
\end{equation}
Here, the uncoupled Hamiltonian $H_0$  describes the energy of the non-interacting QDs,
\begin{equation}
    \label{eq:dipole_tls_ham_terms}
    H_0 = \sum_{j=1}^N \hbar \omega \sigma^+_j \sigma^-_j + h.c.
\end{equation}
while the interaction Hamiltonian $H_I$ describes the DDI between all pairs of QDs
\begin{equation}
    H_I = \sum_{i<j\leq N} J_{ij}\sigma^-_i\sigma^+_j + h.c.
\end{equation}
The coupling strength, $J_{ij}$ between any two dots $i$ and $j$ depends on their general spatial arrangement, as illustrated in~\autoref{fig:dd-interaction}. For dipoles oriented along the x-axis, this can be expressed as:
\begin{equation}
        J_{ij}  = J_0 \frac{\cos(\theta + \alpha)}{d_{ij}^3}.
\end{equation}
In this expression, $d_{ij}$ is the distance between the dots, $\theta$ denotes the angle the connecting vector makes with the x-axis, and $\alpha$ is the angle related to the electric field orientation between them.
The uncoupled Hamiltonian $H_0$ is already diagonal, allowing us to represent its eigenstates in terms of the single QD eigenstates. The total system’s ground state $\ket{b}$ is a state where all QDs are in their respective ground states, i.e., $\ket{g_1, g_2, g_3 \ldots,g_N}$. For $N$ QDs, the first excited manifold consists of $N$ degenerate states, corresponding to only one of the QDs being in the excited state $\ket{e}$ while the other is in state $\ket{g}$. We label these eigenstates as $\ket{a_k} \equiv \ket{g_1, g_2, \ldots e_k, \ldots}$, all these states correspond to the energy eigenstate $E_{\mathrm{GaN}}$.  The interaction Hamiltonian $H_I$ allows mixing of these states, causing the energy spectrum to change. It’s evident from \autoref{eq:dipole_tls_ham_terms} that the interaction doesn't allow mixing with the lowest and highest energy state of uncoupled system, hence they remain unchanged. Upon diagonalizing the total Hamiltonian for the first excited manifold, we obtain the new excitonic energy spectrum, as shown for a 5-QD system in \autoref{fig:energy_levels}.
To understand the dynamics of the system, we are interested in the physical quantities like energy eigenvalues $E_k$, and the transition dipole moment $\mu_{kk'}$ between eigenstates $|k\rangle$ and $|k'\rangle$. To compute $\mu_{kk'}$, we first define the total dipole moment operator $\hat{\mu}_{\mathrm{total}}$ for the system. Considering that each quantum dot exhibits no dipole moment in the ground state and a dipole moment $\mu$ in the excited state, the dipole operator for a single two-level quantum dot can be written as:
\begin{equation}
    \hat{\mu} = \begin{bmatrix}
        0          & \mu_{12} \\
        \mu_{12}^* & \mu      \\
    \end{bmatrix}
\end{equation}
The off-diagonal terms $\mu_{12}$ ($\mu_{12}^*$) represent the transition dipole moments between the ground and excited states of an individual quantum dot. For the entire system of  $N$ quantum dots, the total dipole moment operator is given by:
\begin{equation}
    \hat{\mu}_{\mathrm{total}} =  \sum_{i=1}^{N}\hat{\mu}_i
\end{equation}
The transition dipole moment between two eigenstates is then computed by the matrix element:: $\mu_{kk'}= \langle k| \hat{\mu}_{\mathrm{total}}| k' \rangle$.  The eigenstates $\ket{k}$ and eigenenergies $E_k$ that form the basis for this calculation are obtained by numerically diagonalizing the Hamiltonian.
\subsection{\label{subsec:phonon_photon_couplings}Phononic and Photonic Couplings}
We now examine the transitions between the new energy levels mediated by photonic (illumination) and phononic (lattice vibration) interactions. \footnote{While the comprehensive study of solar radiation effects would be ideal, our focus will be strategically confined to specific radiation frequencies that correspond precisely to the energy gap between targeted levels within our two-dot system, thereby enabling precise control over the transition dynamics.} For our photocell configuration, as presented the the \autoref{fig:energy_levels}, we only consider the ground state and first set of excited levels. Additionally, since the state $\ket{b}$ lacks a dipole moment,  our attention will be directed exclusively toward the dipole moments of the first excited energy manifold. To facilitate efficient transitions, we will  incorporate a pumping field aligned parallel to the dipole moment of the exciton— a condition that can be practically achieved through the use of  a polarizing filter. That said,  as one recalls that the exitonic-photon coupling in QD is given as
\begin{align}
H_{\mathrm{optical}}
& = - \hbar\sum_{\mathbf{k}} \sum_{i=1}^{N} g_{\mathbf{k}}\sigma_x^i  \left( \hat{a}_\mathbf{k} + \hat{a}^{\dagger}_\mathbf{k} \right)
\end{align}
Referring to the Weisskopf-Wigner theory of spontaneous decay, we recognize that the photon induced transition rate between two energy levels is quantitatively described by
\begin{equation}
\label{eq:photon_spon_decay_rate}
\gamma_{\mathrm{photon}}(\omega, \mu) = \frac{\omega^3 |\mu|^2}{3\hbar\pi\epsilon_0 c^3}
\end{equation}
For photonic transitions from the ground state, only states with a non-vanishing transition dipole moment (i.e., bright states) are relevant. Here, $\omega$ is the frequency associated with the transition energy between the two states,  and $\mu$ is the dipole moment. 

To build intuition for the origin of dark states and their role in suppressing radiative recombination, it is instructive to first consider the simple case of two coupled QDs. The interaction splits the excited manifold into a symmetric state $\ket{x_1}$ and and an antisymmetric state $\ket{x_2}$  where
\[
\ket{x_{1}} = \frac{1}{\sqrt{2}}(\ket{a_1} + \ket{a_2}) \quad
\ket{x_{2}} = \frac{1}{\sqrt{2}}(\ket{a_1} - \ket{a_2})
\]
The energy eigenvalues of these new eigenstates are given by $E_{x_{1,2}} = E_{\mathrm{GaN}} \pm J_0$. Now, given that the dipole moment (absolute value) for both the states $\ket{a_{1, 2}}$ is equal to $\abs{\mu}$, we can find the corresponding value for the new eigenstates $\ket{x_{1,2}}$ as below
\begin{equation*}
\label{eq:new-eigenstate-dip-mom}
\begin{gathered}
\abs{\mu_{x_1 \rightarrow b}} = \frac{1}{\sqrt{2}} \abs{\mu_1 + \mu_2} = \sqrt{2}\abs{\mu}\\
\abs{\mu_{x_2 \rightarrow b}} = \frac{1}{\sqrt{2}} \abs{\mu_1 - \mu_2} = 0
\end{gathered}
\end{equation*}
Thus, the symmetric state $\ket{x_1}$ becomes a `bright' state with an enhanced transition dipole moment, while the antisymmetric state $\ket{x_2}$  becomes a `dark' state with a zero transition dipole moment to the ground state. We define the pumping rate from the ground state to this bright state as:
\begin{equation}
\label{eq:pumping-rate}
\gamma_h = \frac{2\omega_{1b}^3{\mu}^2}{3\hbar\pi\epsilon_0 c^3}
\end{equation}
where $\hbar\omega_{1b} = E_{x_1}$ is the energy associated with the pumping transition. We note that the transition rate between the ground state and the symmetric eigenstate is twice that of the uncoupled eigenstates, while transitions rate between the ground state and antisymmetric eigenstate are forbidden, making this state a \emph{dark state}. Since this dark state has no transition dipole moment, all radiative transitions involving this state are forbidden. This principle extends to multi-QD systems, which also exhibit dark states into which radiative recombination is prohibited, as illustrated for the 5-QD system in~\autoref{fig:uncoupled_vs_coupled}. Consequently, non-radiative pathways must be utilized to populate and depopulate these dark states.

In addition to interactions with optical photon, QDs are also coupled to surrounding environment and hence are subject to phononic field modes. We will consider transitions mediated by electron-phonon interactions as the major cause of the decay from bright states to dark states. LO-phonon-electron interactions lead to ultra-fast decays \cite{woggon96} in III-V semiconductors and hence it is reasonable to assume this is much faster than the radiative recombination between bright states and the ground state $\ket{b}$.  We further argue that, due to the large energy difference, a phonon-mediated relaxation to the ground state is much less likely than tunnelling of the electron out of the dot (provided there is an available energy level outside the dot). In our case, this effect is augmented, in part, by the built-in electric field in GaN quantum dots. We will examine this in~\autoref{sec:tunneling_rates}. Phonon-electron interactions have also been a basis of modeling the efficiency enhancements in  light-harvesting antennae complexes~\cite{zhang2016dark, quant_rachet_prx}. The exiton-phonon interaction on optically driven QD  is described by~\cite{Nazir_McCutcheon_2016}
\begin{equation}
	\label{eq:long_exiton_bath_coup}
	H_{\mathrm{vib}} = \sum_{\mathbf{q}} \sum_{i=1}^{N} g_{\mathbf{q}} \sigma_z^i (\hat{b}_{\mathbf{q}} + \hat{b}_{\mathbf{q}}^{\dagger})
\end{equation}
where $\hat{b}_{\mathbf{q}}$ ($\hat{b}_{\mathbf{q}}^{\dagger}$) is the annihilation (creation) operator for the phonons of wavevector $q$ with energy of $\omega_q$ and exciton–phonon coupling $g_q$. It has been suggested that this interaction primarily facilitates transitions within the same energy band since it commutes with $\sum_i \sigma_z^i$~\cite{zhang2016dark, Higgins_Lovett_Gauger_2017}. For the phononic modes, we adopt a structureless Ohmic spectral density, as employed in \cite{quant_rachet_prx}. This choice is motivated by its high-temperature resemblance to the Drude-Lorentz density and its successful application in modeling excitonic transfer in light-harvesting complexes \cite{zhang2016dark}, which subsequently yields the following rates~\cite{quant_rachet_prx, Higgins_Lovett_Gauger_2017}
\begin{equation}
	\label{eq:lomg_transition_rates}
	\gamma_{\mathrm{vib}} (\gamma_x, \omega) = \gamma_x \frac{\omega}{\bar{\omega}_{\mathrm{vib}}}
\end{equation}
where $\gamma_x$ represents a characteristic phonon timescale. $\gamma_x$ would be a free parameter of our simulation, and the parameter range is takes from~\cite{chen2004surface}, and $\bar{\omega}_{vib}$ is the average phonon transition energy specific to the system.

\subsection{\label{sec:tunneling_rates}Charge-extraction mechanism and Tunneling Rates}
Here we present the concept of charge-extraction levels to extract work from the photocells and tunneling rates corresponding to them. We introduce two additional levels aside from the QD energy spectrum, and call these levels \(\ket{\alpha}\) and \(\ket{\beta}\). The electron from the exciton tunnels through the QD barrier into state \(\ket{\alpha}\) where the QD is now left with a net positive charge. The electron now passes through some form of an electrical load at some rate \(\Gamma\), and the system relaxes to another state \(\ket{\beta}\) following which an electron from the bulk semiconductor tunnels into the QD and recombines with the hole to return the whole system to the ground state \(\ket{b}\). The complete energy level scheme for this process is shown in \autoref{fig:uncoupled_vs_coupled}.

The two levels \(\ket{\alpha}\) and \(\ket{\beta}\) can be achieved practically through the conduction band minima (CBM) and the valence band maxima (VBM) of the adjacent n and p type semiconductor regions, respectively. The conduction band minimum (CBM) on the n-doped side is below the excitonic state energy levels. This is by design and tunnelling is allowed from the excitonic levels to the conduction band, since we assume that there are available energy levels above the conduction band minimum. Once the electron has tunnelled out of the quantum dot, it becomes a \emph{hot} electron in the bulk semiconductor, i.e., a carrier with energy higher than the CBM. This hot electron now relaxes to the CBM via the same ultrafast LO-phonon mediated transition as we considered earlier and is discussed in~\cite{woggon96}. Since this transition is much faster than the tunnelling rate, as we will show later, we consider the tunnelling process to be the rate-limiting step in this entire process, and consider the relaxation to the CBM to be nearly instantaneous. Similar arguments apply to the tunnelling of the hole out of the dot to the valence band on the p-doped side. Hence, we consider \(\ket{\alpha}\) to be the CBM on the n-doped bulk semiconductor surrounding the dot and \(\ket{\beta}\) to be the valence band maximum (VBM) on the p-doped bulk semiconductor. 

In the GaN quantum dots, the inbuilt potential profile improves the lateral confinement of the exciton, and the dipole moment of the exciton aligns with the intrinsic electric field. Here, we reasonably assume that the separation of charges happens along the longitudinal dimension of the quantum dot. However, due to this separation of charges, there is a monotonic decrease in the excitonic binding energy as compared to a flat band \cite{cingolani00}, which will aid the tunnelling of electrons in and out of the quantum dots.

Now, since the qualitative theoretical description of the tunneling mechanism is laid out, we will calculate the tunneling rates we are require to carry out numerical calculation for our photocell system. We already have an expression for the incoherent pumping rate due to solar radiation from \autoref{eq:photon_spon_decay_rate} and phononic modes from~\autoref{eq:lomg_transition_rates}. Of primary importance is the tunneling rate of electrons and holes out of the quantum dot. Electron tunneling followed by rapid phonon-mediated relaxation to the CBM corresponds to the transition from a state inside the QD to the CBM state \(\ket{\alpha}\)\footnote{The actual difference between \(\ket{\alpha}\) and \(\ket{\beta}\) is of little consequence because we only care about current and power enhancement, which does not depend on the absolute value of the open-circuit voltage \(V_{OC}\).}. As a simple model, we will evaluate the tunneling rate using the WBK approximation in a method similar to the Gamow theory of alpha decay \cite{gamow28_zur_quant_des_atomk}. The exciton can be treated as loosely bound and, in fact, it is known that the binding energy of the exciton is 2-3 times lower in GaN QDs with a built-in electric field as compared to the flat band case \cite{cingolani00}. Thus the transmission coefficient \(T\) can be approximated as 
\begin{equation}
\label{eq:wkb-tunnel-tran-coeff}
T \approx \exp(-2 \int\limits_a^b dx\frac{\sqrt{2m^{*}_e\abs{V(x)-E}}}{\hbar})
\end{equation}
where \(m^{*}\) is the effective mass, \(E\) is the energy of the single particle and \(V(x)\) is the potential it is tunneling through. \(a\) and \(b\) indicate the classical turning points of the potential. Here, we are also making the assumption that the particle has the same energy on both sides of the barrier. This is true for the average energy in this case since the tunneling is augmented by the in-built electric field and thermal noise in the semiconductor. The tunneling rate is then given by 
\begin{equation}
\label{eq:wkb-tunnel-rate}
\Gamma_{i\alpha} = \nu_cT_{i\alpha}
\end{equation}
where \(\nu_c\) is called the assualt frequency, defined to be equal to the frequency of a classical oscillator with the particle mass and energy oscillating inside the QD. To calculate this for electrons, we consider the classically allowed region inside the QD. Due to band bending, this region is less than the total width of the QD. 
\begin{align}
\begin{split}
\label{eq:assault-freq}
\nu_c &= \frac{v}{2R} \\
&= \frac{1}{2R} \sqrt{\frac{2\pmqty{E^{*}+F_d \mathrm{w}_{\mathrm{d}}/2}}{m_e^{*}}} \\
&= \frac{1}{2\pqty{\frac{E^{*}}{F_d}+\frac{\mathrm{w}_{\mathrm{d}}}{2}}}\sqrt{\frac{2\pmqty{E^{*}+F_d \mathrm{w}_{\mathrm{d}}/2}}{m_e^{*}}}
\end{split}
\end{align}
\(E^{*}\) is the energy of the electron measured from the CBM. With this and the form of the potential known from the parameters given in \cite{DeRinaldis2002,rinaldis04_intrin_elect_field_effec_few}, we can evaluate the tunneling rate for an electron with energy \(E^{*}\)
\begin{gather}
\label{eq:tunnel-rate-expr}
\Gamma_{i\alpha} = \left[2\pqty{\frac{E^{*}}{F_d}+\frac{\mathrm{w}_{\mathrm{d}}}{2}}\right]^{-1} \sqrt{\frac{2\pqty{E^{*}+F_d \mathrm{w}_{\mathrm{d}}/2}}{m_e^{*}}} \\
 \times \exp\left[ \frac{-4\sqrt{2m^{*}_e}}{3\hbar F_d}\left( \pqty{\Delta E_c - F_d \mathrm{w}_{\mathrm{d}}/2 + F_{br}\mathrm{w}_{\mathrm{br}} - E^{*}}^{3/2} \right. \right. \nonumber \\
- \left. \left. \pqty{\Delta E_c - F_d \mathrm{w}_{\mathrm{d}}/2 - E^{*}}^{3/2} \right) \right] \nonumber
\end{gather}
This expression is used to calculate tunneling rates out of the dot and for simplicity, we assume the hole tunneling rate is the same as the electron tunneling rate. 

\section{Quantum Dynamics of the photocell}
The energy level scheme for  uncoupled and coupled system of five GaN quantum dots with aligned dipoles is shown in\footnote{For a system with a nondegenerate Hamiltonian, the application of the secular approximation closes the equations of motion for the populations $p_a(t)$ of the eigenstates $\ket{\omega}$ and decouples them from the coherence dynamics. Since this applies to our system, the energy level population dynamics are modeled using Pauli master equations.} \autoref{fig:uncoupled_vs_coupled}.
\subsection{Uncoupled Model\label{sec:uncoupled_model}}

In case of uncoupled QDs, both quantum dots experience a similar environment in the growth direction, but the interdot spacing is too large to allow dipole-dipole coupling. Since there is no effective interaction between the dots, the system has degenerate energy levels. To model the dynamics of this system, we substitute the interaction picture Hamiltonian into the Liouville-von Neumann equation of motion, including a dissipation term to account for tunneling~\cite{scully1997quantum}.
\begin{equation}
\label{eq:dissipative-eom-2}
\dot{\rho} = \frac{-i}{\hbar} \comm{H_{op}^{\mathrm{int}}}{\rho} - \frac{1}{2}\acomm{\Gamma}{\rho}
\end{equation}
The density operator in the above equation is also in the interaction picture. By tracing over the photonic reservoir~\cite{scully1997quantum} and taking matrix elements of the density operator in the energy eigenbasis and defining $\rho_{ij} \equiv \mel{i}{\rho}{j}$, we obtain the following Pauli master equations for uncoupled QDs corresponding to~\autoref{fig:uncoupled_vs_coupled}(a):
\begin{gather}
\label{eq:pauli_eq_5QD_uncoupled}
    \dot{\rho}_{x_1x_1} = -\bar{\gamma}_h \left[(1+n_{h})\rho_{x_1x_1} - n_{h}\rho_{bb}\right] - \Gamma_{x_1\alpha}\rho_{x_1x_1} \nonumber \\
    \dot{\rho}_{x_2x_2} = -\bar{\gamma}_h \left[(1+n_{h})\rho_{x_2x_2} - n_{h}\rho_{bb}\right] - \Gamma_{x_2\alpha}\rho_{x_2x_2} \nonumber \\
    \dot{\rho}_{x_3x_3} = -\bar{\gamma}_h \left[(1+n_{h})\rho_{x_3x_3} - n_{h}\rho_{bb}\right] - \Gamma_{x_3\alpha}\rho_{x_3x_3} \nonumber \\
    \dot{\rho}_{x_4x_4} = -\bar{\gamma}_h \left[(1+n_{h})\rho_{x_4x_4} - n_{h}\rho_{bb}\right] - \Gamma_{x_4\alpha}\rho_{x_4x_4} \nonumber \\
    \dot{\rho}_{x_5x_5} = -\bar{\gamma}_h \left[(1+n_{h})\rho_{x_5x_5} - n_{h}\rho_{bb}\right] - \Gamma_{x_5\alpha}\rho_{x_5x_5} \nonumber \\
    \dot{\rho}_{\alpha\alpha} = \sum_{i = 1}^5 \Gamma_{x_i\alpha} \rho_{x_i x_i} - (\Gamma+\chi\Gamma)\rho_{\alpha\alpha} \nonumber \\
    \dot{\rho}_{\beta\beta} = \Gamma\rho_{\alpha\alpha} - \Gamma_{\beta b}\rho_{\beta\beta} \nonumber\\
    \sum_{i = 1}^5\rho_{x_i x_i} + \rho_{\alpha\alpha} + \rho_{\beta\beta} + \rho_{bb} = 1  
\end{gather}
Here, we do not re-derive the master equation for the reduced density matrix of a QDs coupled to a photonic bath, given its extensive treatment in prior works~\cite{Dorfman2013, Creatore2013, zhang2016dark, Zhao_2019, Zhang_2015, matthias_2016}. Instead, we employ a phenomenological approach to construct the Pauli master equations, directly referencing the system's energy level diagram. The spontaneous decay term $\Gamma$ is explicitly included, and the anticommutator in \autoref{eq:dissipative-eom-2} simplifies to a single term in the equations, with decay factors manually inserted according to the energy level scheme. Here, \(n_{h}\) is the number of ambient solar photons, $\bar{\gamma}_h \equiv \gamma_h / 2$ as defined in~\autoref{eq:pumping-rate}, \(\Gamma_{x_i\alpha}\) is the electron tunneling rate that takes the photocell system from state \(\ket{x_i}\) (they are degenerate in case of QDs) to \(\ket{\alpha}\), and \(\Gamma_{\beta b}\) is the hole tunneling rate that takes the system from state \(\ket{\beta}\) to \(\ket{b}\). Since all states \(\ket{x_i}\) are identical and exist in identical environments, we consider them equal. Similarly, since the energy levels \(\ket{x_i}\) are degenerate, their spontaneous decay rates will be equal, as will the number of ambient solar photons.

The generalization of the above equations to a system with \(N\) quantum dots is straightforward. For multi-QD system, the Pauli master equations for the unoupled system can be written as, for \(j = (1, 2, \ldots N)\):
\begin{gather}
\label{eq:uncoupled-dot-pme-gen}
\dot{\rho}_{x_jx_j} = -\bar{\gamma}_h \bqty{\pqty{1+n_{h}}\rho_{x_jx_j} - n_{h}\rho_{bb}} - \Gamma_{x_j\alpha}\rho_{x_jx_j} \nonumber \\
\dot{\rho}_{\alpha\alpha} = \sum_{i=1}^{N}\Gamma_{x_i\alpha}\rho_{x_ix_i} - \pqty{\Gamma+\chi\Gamma}\rho_{\alpha\alpha} \nonumber \\
\dot{\rho}_{\beta\beta} = \Gamma\rho_{\alpha\alpha} - \Gamma_{\beta b}\rho_{\beta\beta} \nonumber \\
\rho_{x_1x_1}  +  \ldots + \rho_{x_Nx_N} + \rho_{\alpha\alpha} + \rho_{\beta\beta} + \rho_{bb} = 1
\end{gather}

\subsection{Coupled Model}
\label{sec:org35d74ab}
\begin{figure}[htbp]
\centering
\includegraphics[width=1.0\columnwidth]{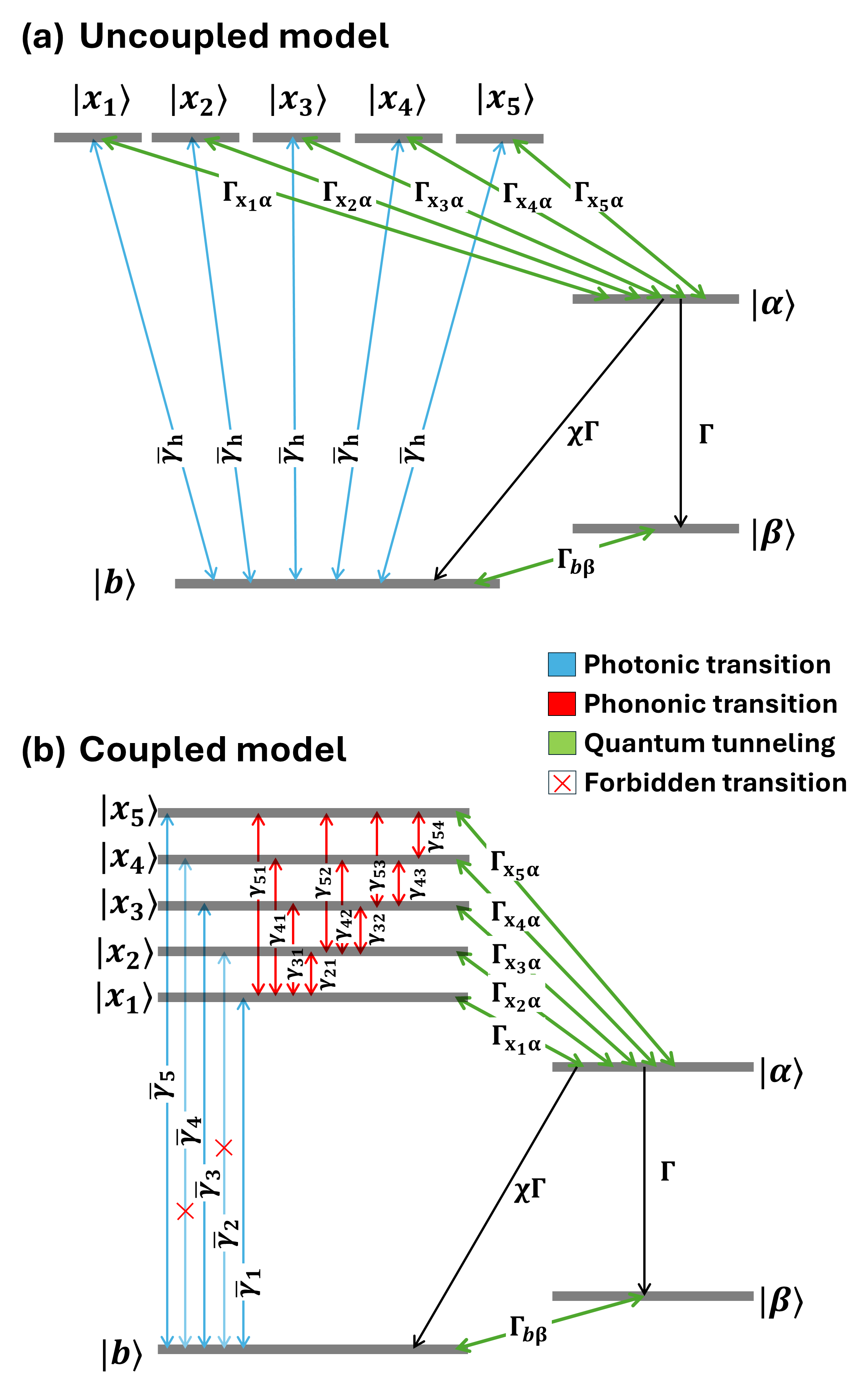}
\caption{Comparison of the energy level schemes for a photocell with (a) uncoupled and (b) dipole-dipole coupled quantum dots (QDs). In the uncoupled case, the first excited manifold consists of degenerate energy levels. When dipole-dipole interaction (DDI) is introduced, this degeneracy is lifted, leading to the formation of a new energy spectrum containing collective bright and dark states. \label{fig:uncoupled_vs_coupled}}
\end{figure}

In the coupled model, we consider dipole-dipole interactions between nearest-neighbour quantum dots only, which can be experimentally realized by tuning the inter-dot separation accordingly. This coupling modifies the excitonic energy spectrum, as detailed in \autoref{sec:excitonic-dipole-dipole-coupling}. We now extend the Pauli master equation framework to incorporate these interactions and derive the population dynamics for the coupled-dot system. The energy level scheme for the photocell incorporating dipole-dipole coupled quantum dots is depicted in \autoref{fig:uncoupled_vs_coupled}(b).

As discussed earlier in~\autoref{sec:excitonic-dipole-dipole-coupling}, photonic transitions are forbidden into dark states due to symmetry constraints, and hence no direct radiative transitions are allowed. The states \(\ket{b}\) and bright states are coupled via an incoherent photonic reservoir, while the states within a energy band are coupled through longitudinal optical (LO) phonons in GaN. Since investigations into the phonon spectrum of GaN quantum dots (QDs) in this specific configuration are still in early stages, we adopt a simplified description of the phonon bath using the Planck distribution of LO-phonon modes. This is a reasonable approximation at room temperature, the regime of interest, where phonon correlations are significantly suppressed, and the bath can be treated as an effectively non-interacting bosonic environment. This justifies the use of the Markovian approximation for the phonon reservoir. To ensure charge extraction from the dark state, the dimensions of the GaN quantum dot are chosen such that the energy of bright states lies above the conduction band minimum (CBM) of the surrounding bulk semiconductor, thereby allowing tunnelling of electrons. Similar constraints are imposed for hole tunnelling from the ground state \(\ket{b}\). Similar to the uncoupled case, the dynamics of the system can be derived by substituting the interaction-picture interaction Hamiltonians into the Liouville–von Neumann equation for the total density matrix. Here, the interaction Hamiltonian includes both the optical and carrier-phonon interactions, i.e. \(\dot{\rho} = \frac{-i}{\hbar} \comm{H_{\mathrm{optical}}^{\mathrm{int}} + H_{\mathrm{vib}}^{\mathrm{int}}}{\rho} - \frac{1}{2}\acomm{\Gamma}{\rho}\)
The pauli master equations describing the population dynamics of coupled 5-QDs with it's energy level diagram illustrated in~\autoref{fig:uncoupled_vs_coupled}(b) is give by
\begin{gather}
\label{eq:pauli_eq_5QD_coupled}
\dot{\rho}_{x_1x_1} = -\sum_{i=2}^5 \gamma_{1i}\left[(1+n(\omega_{1i}))\rho_{x_1x_1} - n(\omega_{1i})\rho_{x_ix_i}\right] \nonumber \\
 - \bar{\gamma}_1\left[(1+n_h)\rho_{x_1x_1} - n_h\rho_{bb}\right] - \Gamma_{x_1\alpha}\rho_{x_1x_1} \nonumber \\
\dot{\rho}_{x_2x_2} = -\sum_{\substack{i=1 \\ i\neq2}}^5 \gamma_{2i}\left[(1+n(\omega_{2i}))\rho_{x_2x_2} - n(\omega_{2i})\rho_{x_ix_i}\right] \nonumber \\
 - \bar{\gamma}_2\left[(1+n_h)\rho_{x_2x_2} - n_h\rho_{bb}\right] - \Gamma_{x_2\alpha}\rho_{x_2x_2} \nonumber \\
\dot{\rho}_{x_3x_3} = -\sum_{\substack{i=1 \\ i\neq3}}^5 \gamma_{3i}\left[(1+n(\omega_{3i}))\rho_{x_3x_3} - n(\omega_{3i})\rho_{x_ix_i}\right] \nonumber \\
 - \bar{\gamma}_3\left[(1+n_h)\rho_{x_3x_3} - n_h\rho_{bb}\right] - \Gamma_{x_3\alpha}\rho_{x_3x_3} \nonumber \\
\dot{\rho}_{x_4x_4} = -\sum_{\substack{i=1 \\ i\neq4}}^5 \gamma_{4i}\left[(1+n(\omega_{4i}))\rho_{x_4x_4} - n(\omega_{4i})\rho_{x_ix_i}\right] \nonumber \\
 - \bar{\gamma}_4\left[(1+n_h)\rho_{x_4x_4} - n_h\rho_{bb}\right] - \Gamma_{x_4\alpha}\rho_{x_4x_4} \nonumber \\
\dot{\rho}_{x_5x_5} = -\sum_{i=1}^4 \gamma_{5i}\left[(1+n(\omega_{5i}))\rho_{x_5x_5} - n(\omega_{5i})\rho_{x_ix_i}\right] \nonumber \\
 - \bar{\gamma}_5\left[(1+n_h)\rho_{x_5x_5} - n_h\rho_{bb}\right] - \Gamma_{x_5\alpha}\rho_{x_5x_5} \nonumber \\
\dot{\rho}_{\alpha\alpha} = \sum_{i = 1}^{5} \Gamma_{x_i \alpha}\rho_{x_i x_i} - (\Gamma+\chi\Gamma)\rho_{\alpha\alpha} \nonumber \\
\dot{\rho}_{\beta\beta} = \Gamma\rho_{\alpha\alpha} - \Gamma_{\beta b}\rho_{\beta\beta} \nonumber \\
\sum_{i = 1}^{5} \Gamma_{x_i \alpha}\rho_{x_i x_i} + \rho_{\alpha\alpha} + \rho_{\beta\beta} + \rho_{bb} = 1
\end{gather}
where \(\gamma_{ij} \equiv \gamma_{\mathrm{vib}}(\gamma_x, \omega_{ij})\) is the LO-phonon-mediated relaxation rate between \(\ket{x_i}\) and \(\ket{x_j}\). Here, we have defined \(\omega_{ij} = \omega_i - \omega_j\). \(n(\omega)\) is the average thermal occupation number of phonons at temperature \(T_a\), given by the Bose-Einstein Distribution:
$n(\omega) = \left[e^{{\hbar \omega}/{k_BT_a}}-1\right]^{-1}$.
Also, \(\bar{\gamma}_i \equiv \gamma_{\mathrm{photon}}(\omega_{ib}, \mu_{ib})\). The values of the relevant quantities, used in calculations are provided in~\autoref{table:param-table}. 
This formulation can be generalized to a system of $N$ quantum dots. Assuming each dot interacts with independent phonon and optical environments, the corresponding generalized Pauli master equations are, for \((j = 1, 2, \ldots, N)\)
\begin{gather}
\label{eq:coupled-dot-pme-gen}
\dot{\rho}_{x_jx_j} = -\sum_{\substack{i=1 \\ i\neq j}}^N \gamma_{ji}\left[(1+n(\omega_{ji}))\rho_{x_jx_j} - n(\omega_{ji})\rho_{x_ix_i}\right] \nonumber \\
 - \bar{\gamma}_j\left[(1+n_h)\rho_{x_jx_j} - n_h\rho_{bb}\right] - \Gamma_{x_j\alpha}\rho_{x_jx_j}, \nonumber \\
\dot{\rho}_{\alpha\alpha} = \sum_{i = 1}^{N} \Gamma_{x_i \alpha}\rho_{x_i x_i} - (\Gamma+\chi\Gamma)\rho_{\alpha\alpha} \nonumber \\
\dot{\rho}_{\beta\beta} = \Gamma\rho_{\alpha\alpha} - \Gamma_{\beta b}\rho_{\beta\beta} \nonumber \\
\sum_{i = 1}^{N}\rho_{x_i x_i} + \rho_{\alpha\alpha} + \rho_{\beta\beta} + \rho_{bb} = 1
\end{gather}
These equations collectively describe the dissipative dynamics of a multi-dot photocell system in the presence of LO-phonon coupling and radiative extraction.
\section{\label{sec:results}Results}
In this section and subsequent subsections, we discuss the key results of our work for various quantum dot configurations and system sizes. We present analyses of the Current-Voltage and Power-Voltage characteristics for systems ranging from 3 to 25 quantum dots (QDs). While our approach can, in principle, be extended to any number of QDs, computational and memory constraints limit our detailed study to systems with up to 25 QDs; beyond this threshold, the calculations become computationally intractable for us. In \autoref{sec:population_dyanmics_curves}, we examine the population dynamics, their saturation behavior, and highlight how our approach contrasts with previous studies. \autoref{sec:PV_IV_characteristic_curves} presents the I-V and P-V characteristic curves, demonstrating a clear advantage of dipole-dipole coupled systems over their uncoupled counterparts. In \autoref{sec:phonon_efficiency_saturation}, we define power enhancement and show its dependence on the phonon transition rate. \autoref{sec:saturation_plot} illustrates the scaling behavior of power enhancement for generic systems, conclusively establishing a positive quantum advantage over classical, uncoupled counterparts even for large system sizes. 

Furthermore, we investigate the dependence of absolute power and power enhancement on parameter space and quantum dot geometry, providing insights into the most practically feasible operational regimes in \autoref{sec:contour_plots}
\begin{table}[hbtp]
	\caption{Values of several parameters used in this work \label{table:param-table}}
	\centering
	\begin{tabular}{lccc}
		\hline
		\textbf{Quantity}                & \textbf{Symbol} & \textbf{Value} & \textbf{Unit}             \\
		\hline
		GaN Bandgap                      & \(E_{\mathrm{GaN}}\)         & 3.51           & \si{\electronvolt}        \\
            \hline
    	Charge Extraction Level Gap                     & \(E_{\alpha\beta}\)         & 3.42           & \si{\electronvolt}        \\
		\hline
		Conduction band dislocation      & \(\Delta E_c\)  & 0.0            & \si{\electronvolt}        \\
		\hline
		Valence band dislocation         & \(\Delta E_v\)  & 0.0            & \si{\electronvolt}        \\
		\hline
		Electron effective mass          & \(m^{*}_e\)     & 0.2            & $m_{\mathrm{e}}$        \\
		\hline
		Hole effective mass              & \(m^{*}_h\)     & 1.0            & $m_{\mathrm{e}}$        \\
		\hline
		Relative permittivity            & \(\epsilon\)    & 9.6            &                           \\
		\hline
		Characteristic phonon rate                & \(\gamma_x\)         & 0.1            & \si{\per\femto\second}          \\ 
        \hline
        Quantum dot width                & \(\mathrm{w}_{\mathrm{d}}\)         & 2.7            & \si{\nano\metre}          \\
		\hline
         Exitonic Dipole Moment               & \(\mu\)         & $0.8\mathrm{w}_{\mathrm{d}}$            & e $\cdot$ \si{\nano\metre}          \\
		\hline
		QD intrinsic electric field      & \(F_d\)         & 0.54           & \si{\volt\per\nano\metre} \\
		\hline
		Barrier intrinsic electric field & \(F_{br}\)      & 0.57           & \si{\volt\per\nano\metre} \\
		\hline
		Radiative recombination rate     & \(\chi\)        & 0.20           &                           \\
		\hline
		Ambient phonon temperature       & \(T_a\)         & 300            & \si{\kelvin}              \\
		\hline
		Number of solar photons          & \(n_h\)         & 60000          &                           \\
		\hline
	\end{tabular}
\end{table}
\vspace{1cm}

\subsection{\label{sec:population_dyanmics_curves}Population Dynamics}
The dynamics of both the coupled and uncoupled photocell models are described by the Pauli master equations, \autoref{eq:pauli_eq_5QD_uncoupled} and \autoref{eq:pauli_eq_5QD_coupled}, respectively. As these systems of differential equations can be stiff for certain parameter regimes, we have employed the RADAU implicit Runge-Kutta solver to obtain robust numerical solutions. The simulations are based on the material and system parameters for GaN QDs listed in \autoref{table:param-table}. These values are used throughout our analysis unless explicitly stated otherwise. Any additional parameters specific to a particular result will be mentioned alongside its presentation.

\begin{figure}[htbp]
\centering
\includegraphics[width=1.0\columnwidth]{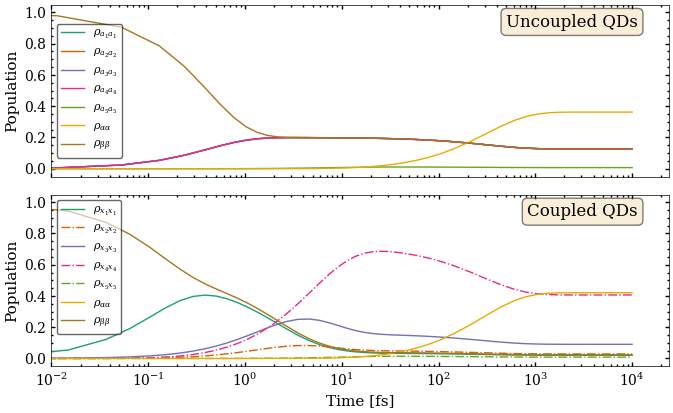}
\caption{Population dynamics of coupled and uncoupled photocells for a 5-QD system at 300K. The dotted curves highlight the significant population of the dark states in the coupled quantum dot (QD) system. The logarithmic time axis provides a clear view of the saturation dynamics.\label{fig:population-dynamics}}
\end{figure}

To analyze the system's behavior, the Pauli master equations were solved numerically until the populations reached a steady state. As a representative example, the population dynamics for a 5-QD system in a (1, 1, 5) configuration are presented in~\autoref{fig:population-dynamics}, using specific rate parameters of \(\gamma_x = \SI{0.1}{\femto\second}^{-1}\) and \(\Gamma = 0.08\); all other parameters are as given in~\autoref{table:param-table}. A key feature observed in the results is the significant population build-up in the dark states for the coupled system, which directly illustrates the exciton-trapping due to delocalization, the mechanism central to our model.

It is important to contrast our approach with related prior work, such as that by Svidzinsky, Dorfman, and Creatore \cite{Svidzinsky2011, Dorfman2013, Creatore2013}. The notable differences in our results are expected and arise from fundamental model distinctions. Whereas those studies focused on quantum heat engines in molecular systems with phonon-mediated charge transfer, our work models a completely solid-state device where charge separation is governed by electron tunneling. This reliance on tunneling makes our system's power output highly sensitive to barrier-dependent tunneling rates but also distinguishes its operational principles. For instance, the unidirectional nature of the tunneling transitions prevents lasing-like behavior. Furthermore, our model primarily considers the strong dipole-dipole coupling and the effects of the intrinsic electric field, neglecting Fano-Agarwal interference terms, which are considered a much weaker effect in this context. Additionally, we use a practically similar photocell device architecture based on a quantum dot-based solid-state system. 

\subsection{\label{sec:PV_IV_characteristic_curves}Current-Voltage (I-V) and Power-Voltage (P-V) Characteristics}
\begin{figure}[htbp]
\centering
\includegraphics[width=1.0\columnwidth]{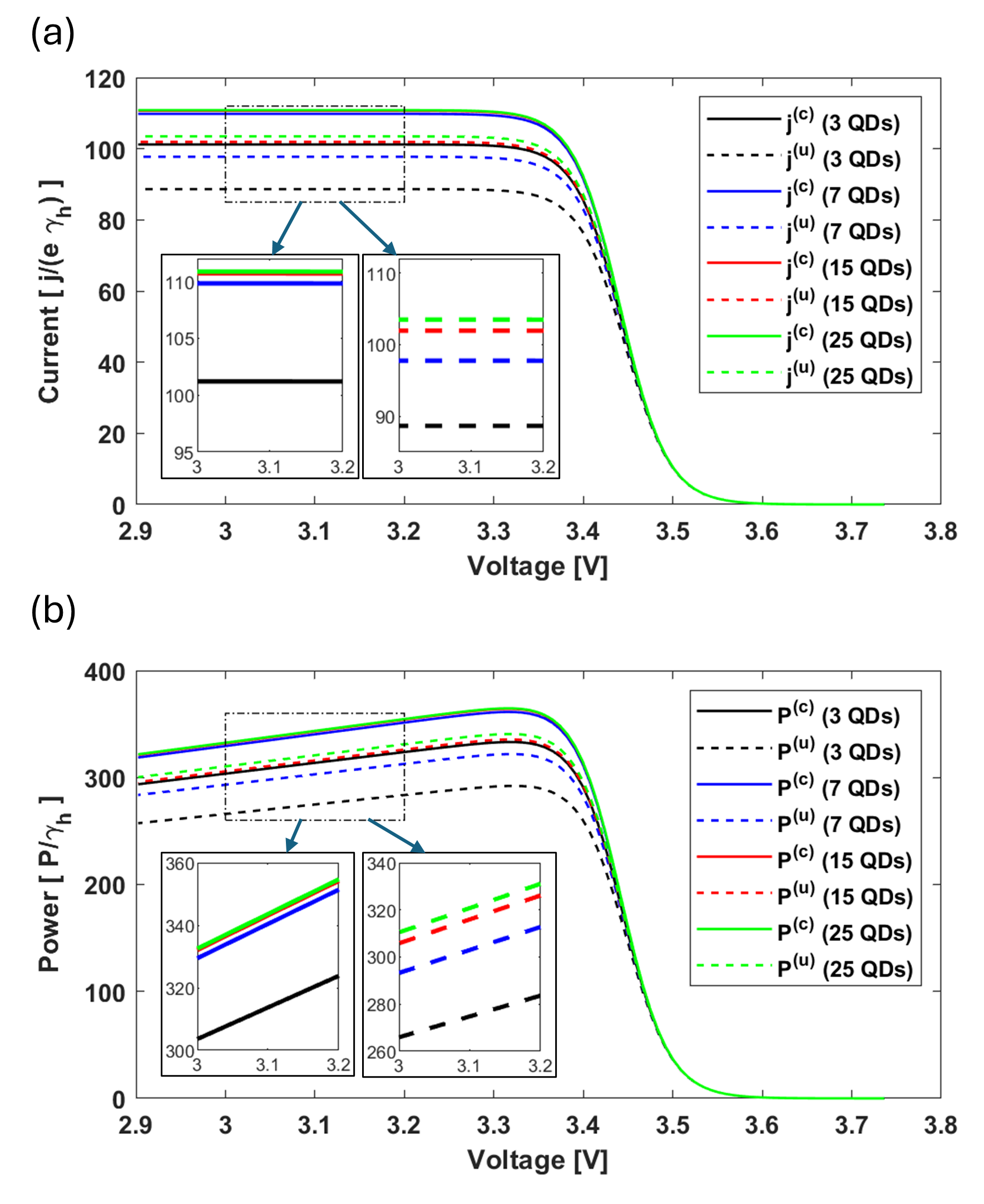}
\caption{Current-voltage (I-V) and Power-Voltage (P-V) characteristics for coupled $^{(c)}$ and uncoupled $^{(u)}$ photocells across varying numbers of QDs are shown in the labels. The plots clearly demonstrate current and power enhancement in the coupled systems, with the improvement saturating as $N$ increases. This is reflected by the plateauing distance between the $j^{(c)}$ and $j^{(u)}$ curves. Notably, the observed power enhancement primarily arises from the increase in current.}
\label{fig:curr-pow-enha}
\end{figure}
Next, we analyze the I-V and P-V characteristics to quantify the performance enhancement across systems with a varying number of QDs, as shown in~\autoref{fig:curr-pow-enha}. The external load on the photocell is varied by adjusting the parameter \(\Gamma\). The voltage across the photocell terminals is obtained from the steady-state populations using the relation\cite{Scully2010}:
\begin{equation}
\label{eq:voltage-expr}
eV = E_{\alpha\beta}  + k_bT_a \ln(\rho_{\alpha\alpha}/\rho_{\beta\beta})
\end{equation}
The limits \(\Gamma \rightarrow 0\) corresponds to the open-circuit condition \(j\to 0\), while \(V = 0\) represents the short-circuit regime. The steady-state current, defined as \(j \equiv e\Gamma \rho_{\alpha\alpha}\), and the output power, \(P=j \cdot V\), are computed from the steady-state solutions of the master equations.

A key finding is that both the photocurrent and the output power saturate as the number of quantum dots in the system increases. This saturation in current, which we will discuss further in~\autoref{sec:saturation_plot}, while conclusively establishing an advantage over purely classical photocell systems, is attributed to a bottleneck in the charge extraction rates that emerges as the system size grows, leading to the observed plateau in performance. From these characteristics, we conclude that the power enhancement in our system primarily manifests as current enhancement, since a significantly higher current is drawn from the coupled photocell compared to the uncoupled one at the same operating voltage. Crucially, the insets in~\autoref{fig:curr-pow-enha} provide visual confirmation of this trend. The clear separation between the solid curves (coupled) and the dotted curves (uncoupled) represents the performance gap, which persists as the number of QDs increases. This indicates that the positive efficiency enhancement saturates with increasing system size rather than vanishing - a confirmation of advantage over purely classical counterparts; this observation is discussed further in~\autoref{sec:saturation_plot}.
\subsection{\label{sec:phonon_efficiency_saturation}Optimization of Phonon-Mediated Relaxation Rate}
\begin{figure}[htbp]
\centering
\includegraphics[width=1.0\columnwidth]{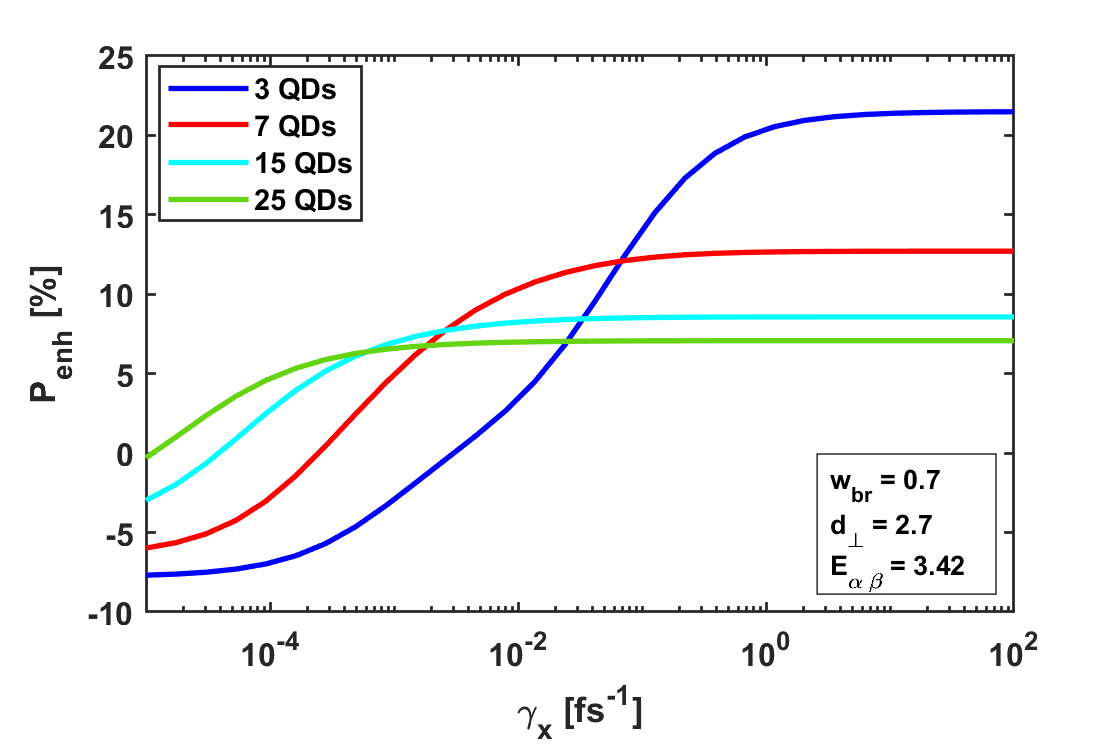}
\caption{Relative power enhancement (\(P_{\mathrm{enh}}[\%]\)) as a function of the phonon-mediated relaxation rate \((\gamma_x)\) to the dark state. The enhancement increases with the rate before reaching a saturation point.\label{fig:pow-enha-vary-phonon-rate_new}}
\end{figure}
The phonon-mediated transition rate, \(\gamma_x\), between the bright and dark states is another critical parameter influencing the photocell's efficiency. To identify the optimal operational regime, we evaluate the peak relative power enhancement, defined as:
\begin{equation}
\label{eq:rel-pow-eff-expr}
P_{\mathrm{enh}} = \frac{P^{c}_{\mathrm{max}} - P^{u}_{\mathrm{max}}}{P^{u}_{\mathrm{max}}}
\end{equation}
The dependence of \(P_{\mathrm{enh}}\) on \(\gamma_x\) is illustrated in~\autoref{fig:pow-enha-vary-phonon-rate_new}. This behavior is expected; if the phonon-mediated transition is too weak, the excited bright states cannot be depleted into the symmetry-protected dark states fast enough to effectively prevent radiative recombination. The enhancement increases with \(\gamma_x\) until it saturates. This saturation occurs because once the phonon relaxation is no longer the rate-limiting step in the population transfer process, further increasing its speed yields no additional benefit.
To ensure operation in the optimal regime, we select the highest feasible value of \(\gamma_x\) permitted by the Markovian dynamics framework employed in our model. This approach maximizes the efficiency enhancement while maintaining the validity of our theoretical assumptions.
\subsection{\label{sec:saturation_plot}Power Enhancement Scaling Behavior with System Size}
\begin{figure}[!htbp]
    \centering
    \includegraphics[width=1\linewidth]{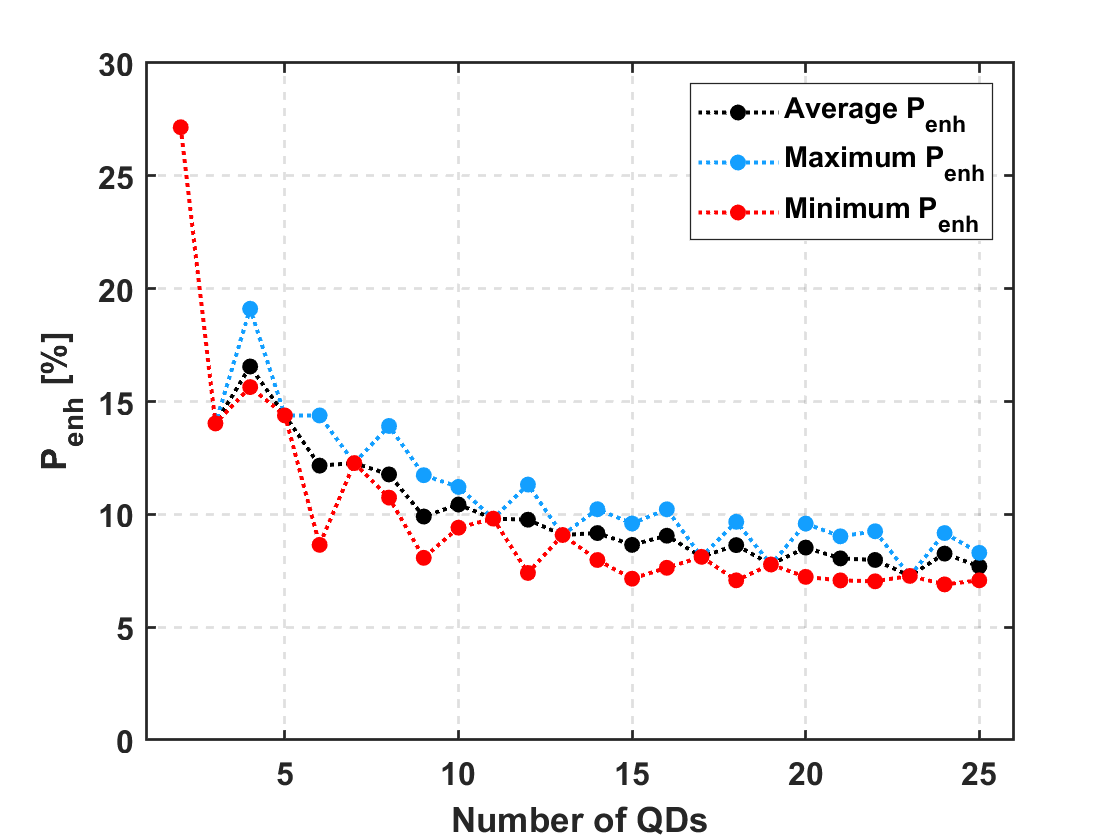}
    \caption{Average $P_{\mathrm{enh}} \%$ versus the number of quantum dots (QDs). : The red line indicates the maximum efficiency enhancement observed across all QD configurations for a specific QD count, while the green line denotes the corresponding minimum enhancement. The blue curve represents the average enhancement across all configurations for that given number of QDs.}
    \label{fig:effi_saturation}
\end{figure}
To assess the viability of this enhancement mechanism for larger, practical devices, we analyzed the performance as a function of the number of quantum dots (N) in the system. \autoref{fig:effi_saturation} illustrates the system's efficiency as a function of the number of quantum dots (QDs). We present the maximum, minimum, and average efficiency enhancement observed across various QD configurations for a specific QD count. A notable observation is that the average enhancement does not diminish to zero for large number of QDs. Instead, it appears to saturate at a significant positive value of approximately $\sim 7\%$, indicating that the coupled quantum dot system provides a clear overall advantage.

\begin{figure*}[hbtp]
	\centering
	\includegraphics[width=1.0\linewidth]{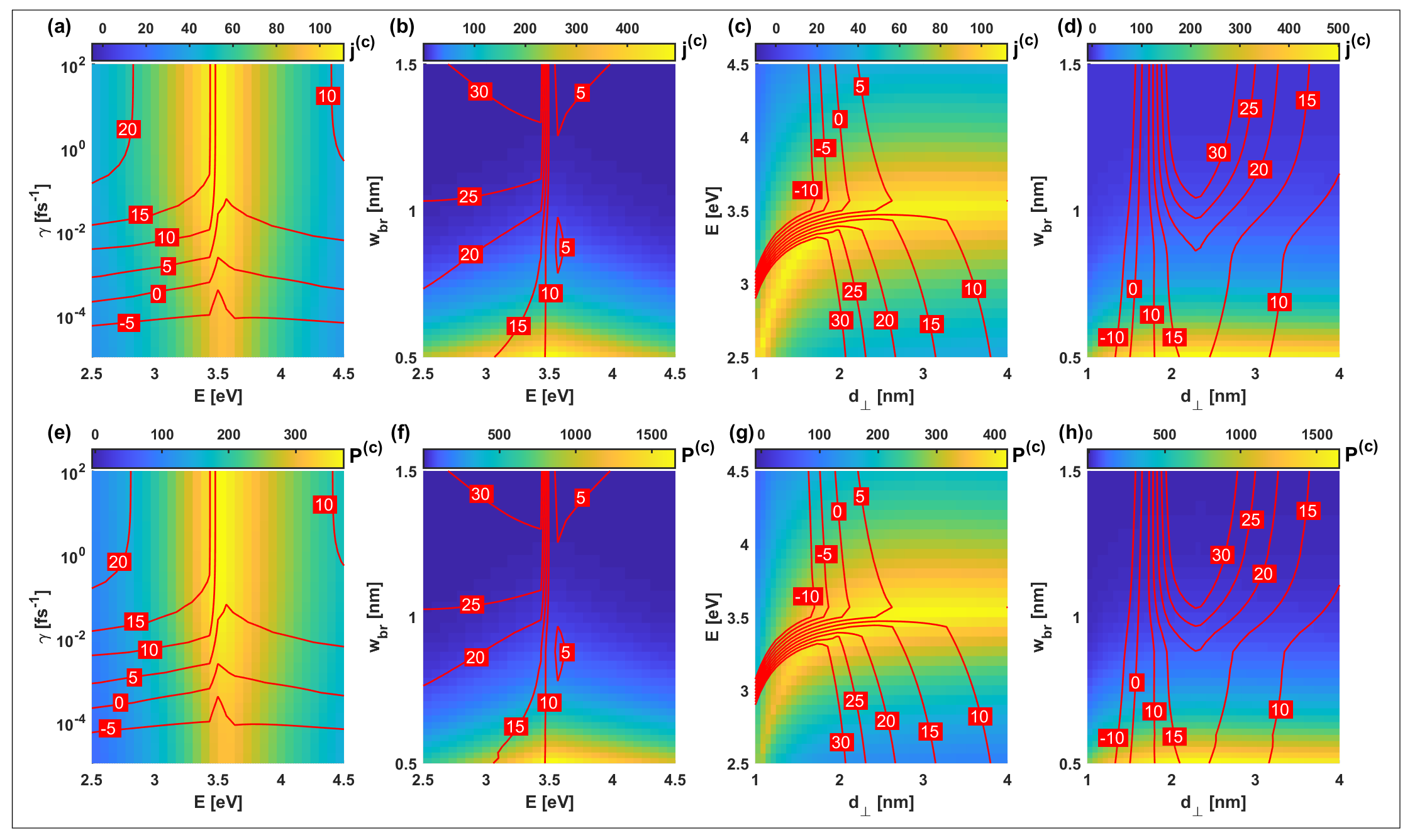}
	\caption{Parameter space exploration for the (1, 5, 1) quantum dot configuration, illustrating the dependence on key physical parameters. The top row (a-d) shows the absolute current in the coupled case ($j^c$) with relative current enhancement shown as contour lines. The bottom row (e-h) shows the corresponding absolute power output of the coupled system ($P^c$) with relative power enhancement as contour lines. Each column pair investigates a different set of parameters: (a, e) Phonon-mediated relaxation rate ($\gamma_x$) versus the charge extraction energy gap ($E_{\alpha\beta}$), demonstrating that enhancement saturates for high $\gamma_x$. (b, f) Barrier width ($\mathrm{w}_{\mathrm{br}}$) versus $E_{\alpha\beta}$, highlighting that practical power output requires small barrier widths ($\mathrm{w}_{\mathrm{br}} \lesssim 1.2$ nm) for efficient tunneling. (c, g) Interdot distance ($d_\perp$) versus $E_{\alpha\beta}$, identifying an optimal regime where $d_\perp \gtrsim 2.0$ nm for meaningful enhancement. (d, h) The primary geometric parameters, $d_\perp$ versus $\mathrm{w}_{\mathrm{br}}$, revealing the optimal design space for maximizing both power output and relative enhancement.}
\label{fig:parameter_space_contour}
\end{figure*}

While the precise physical mechanisms underlying this saturation require further detailed investigation, we can propose a strong hypothesis based on the system's kinetics. This saturation behavior is analogous to the plateau observed in \autoref{fig:pow-enha-vary-phonon-rate_new}, where efficiency enhancement plateaued despite increasing photonic transition rates. In that case, the exciton population transfer rate appeared to reach a threshold limit, suggesting the presence of a rate-limiting step in the population transfer process.
Similarly, we hypothesize that a comparable transport limitation may govern the present system's behavior. Specifically, we propose that the tunneling rate from the quantum dots to the charge extraction levels may represent a bottleneck that becomes increasingly pronounced as the QD count increases. Under this hypothesis, while additional quantum dots contribute to light absorption and suppressed radiative recombination leading to improved exciton generation, the rate at which these excitons can be extracted from the QD array becomes the dominant factor limiting overall efficiency. This shift in the dominant limiting factor could explain the observed efficiency plateau. Further investigation is warranted to rigorously explore and confirm this proposed mechanism.
 
\section{\label{sec:contour_plots}Contour Plots for 6 Level System: (1, 5, 1) Config}

We now present the dependence of absolute power and current values for coupled QDs in the mentioned configuration and their relative enhancements defined according to~\autoref{eq:rel-pow-eff-expr}.
\paragraph{Simultaneous dependence on $\mathrm{w}_{\mathrm{br}}$ and $E_{\alpha\beta}$:}
As depicted in \autoref{fig:parameter_space_contour}(b, f), the absolute power, currents, and their relative enhancements for coupled ($P^{c}$) cases exhibit a direct correlation with both the energy difference between load energy levels ($E_{\alpha\beta}$) and the barrier width ($\mathrm{w}{\mathrm{br}}$). Practical values of absolute power require $\mathrm{w}{\mathrm{br}} \lesssim 1.2$ nm, which stems from the exponential suppression of tunneling rates proportional to $\exp{-2 \int\limits_0^{\mathrm{w}_{\mathrm{br}}} \frac{\sqrt{2m^{*}e\abs{V-E}}} {\hbar}dx}$. Consequently, the population of the $\ket{\alpha}$ state ($\rho_{\alpha\alpha}$) becomes significantly reduced, leading to suppressed steady-state currents $j = e\Gamma \rho{\alpha \alpha}$.
The presented results are based on a fixed value of $d_\perp = 2.7\, \si{nm}$ , though this trend remains consistent across the full range of $d_\perp$ values. Since population dynamics reveal that slow tunneling rates constitute the primary limitation for efficient photocell operation, identifying optimal parameter regimes for both geometry and load becomes crucial for maximizing system efficiency. The controllable parameters are interdot spacing ($d_{\perp}$), barrier width ($\mathrm{w}{\mathrm{br}}$), and load energies ($E{\alpha\beta}$).

\paragraph{Dependence on $d_\perp$ and $E_{\alpha\beta}$:} \autoref{fig:parameter_space_contour}(c, g) demonstrates how absolute power, current enhancements, and their relative values depend on the energy difference between load levels ($E_{\alpha\beta}$) and interdot spacing ($d_\perp$). Throughout the entire $d_\perp$ range and within $2 \, \si{eV}$  $\lesssim E_{\alpha\beta} \lesssim 5 \, \si{eV}$, absolute power values remain practically significant. However, achieving meaningful relative power enhancement requires $d_\perp \gtrsim 2.0 \, \si{nm}$. Given that the Pauli master equation governs our system's dynamics, larger values of \(d_\perp\) are preferable as they better align with the weak coupling regime. Furthermore, for larger \(d_\perp\) values, particularly when \(E_{\alpha\beta} \approx E_{\mathrm{GaN}}\), tunneling rates are enhanced due to an increase in the energy difference term $E$ in \autoref{eq:wkb-tunnel-tran-coeff}) and \autoref{eq:assault-freq}. This improvement facilitates tunneling to the \(\ket{\alpha}\) level, thereby increasing the steady-state current and, consequently, the power output.

\paragraph{Dependence on $\gamma_x$ and $E_{\alpha\beta}$}
\autoref{fig:parameter_space_contour}(a, e) illustrates the relationship between $E_{\alpha\beta}$ and $\gamma_x$. As expected, higher phononic transition rates promote population transfer into dark states, reducing radiative recombination. However, these effects saturate beyond a specific threshold. The optimal $E_{\alpha\beta}$ range identified here aligns with previous analyses, further validating our findings.

\paragraph{Power dependence on QD geometry} \autoref{fig:parameter_space_contour}(d, h) shows how absolute power, currents, and relative enhancements for coupled ($P^c$) cases depend on barrier width ($\mathrm{w}_{\mathrm{br}}$) and interdot spacing ($d_\perp$). For fixed $E_{\alpha\beta}$ and cold transition rate $\gamma_x = 0.1$ fs$^{-1}$, peak enhancements in practically feasible power output regions exhibit positive relative enhancement. The optimal parameter space encompasses $d_\perp$ between $2-3~\si{nm}$, while $\mathrm{w}_{\mathrm{br}}$ should be minimized within fabrication constraints.
\section{\label{sec:conclusion} Discussion and Conclusion}
In this paper, we propose a quantum-enhanced photocell based on GaN quantum dots and examine its current and power enhancement over its purely classical counterpart. We establish that by exploiting the strong intrinsic electric fields in GaN and the quantum coherence arising from excitonic dipole-dipole interaction, we can create dark states that suppress radiative recombination and breaks the detailed balance. We have systematically studied the proposed device architechture for various configurations of sandwiched quantum dots for upto 25 QDs, and studied excitonic dipole-dipole coupling in these configurations of GaN quantum dots to understand the performance enhancement over its classical (uncoupled) counterpart. The core of this enhancement lies in the formation of optically dark excitonic states and population transfer facilitated by phononic transitions. As shown by the population dynamics in \autoref{fig:population-dynamics}, these dark states act as an effective ``trap" or reservoir for excitons, prohibiting them from rapid radiative recombination hence breaking the principle of detailed balance, which manifests as a enhancement in the steady-state photocurrent, and consequently, the power output of the device, as evidenced by the I-V and P-V characteristics in \autoref{fig:curr-pow-enha}. 

Our numerical simulations, based on a Pauli Master Equation formalism, confirm these performance gains across systems of various sizes. We have established the critical roles of key parameters, including phonon-mediated relaxation rates and the geometric layout of the QDs, and identified optimal operational regimes through extensive parameter sweeps as presented in \autoref{sec:contour_plots}. Since the charge extraction process, which relies on quantum tunneling, is highly sensitive to the geometric parameters of the array. The results from our parameter space exploration in \autoref{fig:parameter_space_contour} highlight the critical trade-off: the interdot spacing $d_\perp$ must be small enough to ensure dipole dipole coupling, while the barrier width $\mathrm{w}_{\mathrm{br}}$ must be minimized to allow for efficient tunneling.  This high sensitivity to nanoscale dimensions underscores the significant fabrication challenges that would need to be addressed to realize such a device experimentally. Furthermore, our investigation into device scalability shows that quantum enhancement persists in larger systems, saturating at a positive value. The observation that average power enhancement remains at approximately ($\sim7\%$) for systems with up to 25 QDs (\autoref{fig:effi_saturation}) is a key finding. This indicates that the quantum advantage is maintained as system size increases, which is critical for practical applications. Our work also suggests that future optimization for larger systems should focus not only on quantum coupling but also on the architecture of charge transport pathways.

Our work establishes a viable pathway for leveraging collective quantum effects in energy-harvesting devices. The principles outlined here offer a promising direction for the design of next-generation, high-efficiency photovoltaic technologies that could push beyond the conventional Shockley-Queisser limit. Future work should focus on experimental validation of the proposed DDI mechanism in GaN QD arrays and on developing more sophisticated models that incorporate effects of material disorder and temperature to better guide the practical realization of these devices.

\begin{acknowledgments}
This work is supported by the Anusandhan National Research Foundation (ANRF), India, via grant no. CRG/2021/007060. The authors thank the Ministry of Electronics and Information Technology (MeitY) and the National Quantum Mission (NQM) of DST, India, for supporting A.D.'s research through grant nos. 4(3)/2024-ITEA and DST/QTC/NQM/QC/2024/1. We also acknowledge the National Supercomputing Mission (NSM) for providing computing resources of ‘PARAM Ganga’ at IIT Roorkee, which is implemented by C-DAC and supported by the Ministry of Electronics and Information Technology (MeitY) and Department of Science and Technology (DST), Government of India. 
\end{acknowledgments}

\bibliography{main}

\end{document}